# DETECTING NONLINEARITY IN PRESSURE DATA INSIDE THE DRAFT TUBE OF A REAL FRANCIS TURBINE

Stefano Sello[1]

ABSTRACT

A general method for testing *nonlinearity* in time series is described and applied to measurements of different pressure data inside the draft tube surge of a real Francis turbine. Comparing the current original time series to an ensemble of *surrogates* time series, suitably constructed to *mimic* the linear properties of the original one, we was able to distinguish a linear stochastic from a nonlinear deterministic behaviour and, moreover, to quantify the degree of nonlinearity present in the related dynamics. The problem of detecting nonlinear structure in real data is quite complicated by the influence of various contaminations, like broadband noise and/or long coherence times. These difficulties have been overcame using the combination of a suitable *nonlinear filtering* technique and a qualitative *redundancy* statistic analysis. The above investigations allow a quantitative characterization of different dynamical regimes of motion of gas cavities inside real turbines and, moreover, allow to support the reliability of some related mathematical modelizations.

[1]CISE- Tecnologie Innovative, Applied Mathematics Section
Via Reggio Emilia 39 - 20090 Segrate MILANO (Italy)
Tel. 39+2+21672216  Fax 39+2+21672620  E-mail 284allse@alliant.cise.it

## 1. INTRODUCTION

The analysis of physical quantities from experimental measurements, performed on real dynamical systems, often assume hypotheses that can be confirmed only *a posteriori*, following the interpretative stage. This is a common situation when we move in a strictly linear framework, like the classical Fourier analysis. In fact, irregular and broadband signals, generic in nonlinear dynamical systems, have been discarded in the past as unwelcome "noise" or random information. The characterization of these complex signals and the extraction of physically interesting and useful features, is the main goal of the new methods for the analysis of real time series. Much of the current interest in nonlinear signal analysis arises mainly from the recognition that some complex nonlinear deterministic motions, called *chaos* for simplicity, play a significant role in physical systems. This fact has led to completely new techniques for time series analysis and consequently new ways to understand and to appreciate important, i.e. physically significant, aspects of experimental data. Nonlinearity or nonlinear determinism is a necessary condition for chaotic behaviour and, more generally, is a remarkable feature of every real process.

The problem of deducing the dynamics of a given physical system from measured data is a well known challenge for experimental analysts. The new concepts and techniques involved in time series analysis are in the framework of nonlinear dynamics and theory of deterministic chaos. In the next section we illustrate some remarks about the analysis of time series with a particular emphasis on the detection of the underlying nonlinearity, and the problem of his reliable quantitative evaluation. Furthermore, we will describe a practical application of these concepts and techniques, on different time series generated from pressure measurements inside the draft tube surge of the real Francis turbine installed in the *NOVE Hydroelectric Power Plant of ENEL SpA (ITALY)*. The identification and quantification of the related underlying nonlinear deterministic/chaotic dynamics, allows a deeper insight of the features of this real physical system and, moreover, it gives useful information about some possible mathematical models suggested in previous works [1],[2].

## 2. TIME SERIES ANALYSIS: DETECTING NONLINEARITY

Since the time of Poincaré it has been recognized that even simple, but nonlinear, physical systems can produce very complicated dynamical behaviours. This work mainly deals with the problem of detecting nonlinearity in real time series; therefore it is convenient to spent some words about the concept of nonlinearity associated to time series.

Let $x(t)$ a given time series of $N$ values taken at regular intervals of time: $t=t_0, t_1=t_0+dt,...,t_{N-1}=t_0+(N-1)dt$, where dt is the constant sampling interval between successive measurements. Qualitatively, if the relation between two generic successive values, $x(t)$ and $x(t+dt)$, is expressed by a nonlinear function, we can say that the time series is nonlinear. A more rigorous definition, related with the practical methods for detection of nonlinearity, consists in the proof of an inconsistency with a *linear stochastic process.* This approach, common in statistical analysis, is a proper test of data against a null hypothesis, here a linear stochastic model [3]. The main goal of the analysis, is to test whether all the dependencies in a given time series can be explained on the linear level and moreover evaluate, if any, all the quantitative differences between the original data under study and a proper set of realizations of a linear stochastic process with the *same linear properties.* If we detect some significant inconsistency with the assumed null hypothesis, we will consider the time series to be *nonlinear*. We note that the detection of nonlinearity is a first step in a search for chaotic behaviour, and for that reason many authors, working on applied nonlinear dynamics, have recently proposed different approaches to this problem [4],[5].

A statistical test requires a well defined mathematical procedures for testing a given null hypothesis (here a linear stochastic process) by evaluating some statistical quantity, the value of which discriminates the confidence interval to reject it. Theiler et al., in 1992, suggest a way to a quantitative statistic using the concept of a *"surrogate data"* [6]. The basic idea is to artificially generate a set of time series which *mimic* some of the features of the original data except the properties which we are testing for. In the case of testing for nonlinearity the surrogate data should have the same Fourier spectrum and autocorrelation function, i.e. the *same* linear properties as the original time series under study. For that reason these surrogates are generated as realizations of a linear stochastic process. The method starts applying a *discrete Fourier transform operator*, $\mathscr{F}$, to the original time series:

$$X_{\mathscr{F}}(\omega) = \mathscr{F}\{x(t)\} = \sum_{k=0}^{N-1} x(t_k) e^{2\pi i \omega k dt} \tag{1}$$

This complex Fourier transform we rewrite as:

$$X_{\mathscr{F}}(\omega) = A(\omega) \, e^{i\Phi(\omega)}$$

where $A(\omega)$ and $\Phi(\omega)$ are the amplitude and the phase respectively. To produce a realization of a linear stochastic process, we *randomize* the phases at each frequency sampled by an independent random variable, $\rho(\omega)$, uniformly distributed in the interval $[0, 2\pi)$:

$$\tilde{X}_{\mathscr{F}}(\omega) = A(\omega) \, e^{i[\Phi(\omega) + \rho(\omega)]} \tag{2}$$

In this way the original Fourier spectrum, i.e. the absolute values of the original Fourier coefficients, are invariant quantities. The generic surrogate time series is finally given by the application of an inverse Fourier transform to obtain:

$$\tilde{x}(t) = \mathscr{F}^{-1}\{\tilde{X}_{\mathscr{F}}(\omega)\} = \mathscr{F}^{-1}\{X_{\mathscr{F}}(\omega) e^{i\rho(\omega)}\} \tag{3}$$

By construction, the artificial time series (3) is a particular realization of a linear stochastic process with the same spectrum and, by the Weiner-Khintchine theorem, with the same autocorrelation function as the original data [5]. In our case of testing for nonlinearity, we need a proper ensemble of surrogate time series (3) to perform a reliable statistic through some quantity, Q, able to discriminating nonlinearity. In principle, any nonlinear statistic can be used, as documented in literature [8],[9]. In this work we used the *Takens best estimator of correlation dimension* [7]:

$$Q = D_T := \frac{C(r_0)}{\int_0^{r_0} \frac{C(r)}{r} dr} \tag{4}$$

where $r_0$ is an upper cut-off, and $C(r)$ is the correlation integral derived from the Grassberger-Procaccia algorithm, [10], in the extended version of Theiler [11]:

$$C_m(\alpha, N, r) = \frac{2}{(N+1-\alpha)(N-\alpha)} \sum_{j=\alpha}^{N-1} \sum_{i=0}^{N-1-j} \theta(r - \|\underline{y}(t_i) - \underline{y}(t_{i+j})\|) \tag{5}$$

where $\underline{y}$ is a vector in a proper reconstructed embedding space of dimension $m$ that preserves all the invariant characteristics of sets in the original phase space [12],[13]. The *embedding vectors* in (5) are generated starting from a scalar quantity, $x$, using the *delay time technique*:

$$\underline{y}(t) = (x(t), x(t+\tau), ..., x(t+(m-1)\tau))^T \tag{6}$$

where $\tau$ is the *delay-time*, i.e. the time interval between successive elements, and it is chosen such that the vector contains a maximum amount of information, without the components becoming completely uncorrelated. The discriminating statistic, eq.(4), is then computed for each of the surrogate time series and for the original one. If the numerical value obtained for the original data is *significantly different* from those obtained for the ensemble of surrogate data, then the null hypothesis of a linear stochastic process can be rejected and the original data classified as nonlinear. As a measure of *significance*, i.e. how significantly different the original data is from the ensemble of surrogates, we use the *number of sigmas* (or standard deviations) defined by:

$$S = \frac{|Q - \langle Q_{surr} \rangle|}{\sigma_{surr}} \quad (7)$$

where $Q$ is the discriminating statistic computed for the original data, $\langle Q_{surr} \rangle$ and $\sigma_{surr}$ are respectively the mean value and the standard deviation, computed for the surrogates.

Formally, the above method provides a measure of statistical confidence, in term of probability, that the null hypothesis is false [14]. More precisely, the null hypothesis (here a linear stochastic process) is rejected, and thus the result is considered significant, if the *residual* probability, *p*, of the above hypothesis is lower than a chosen critical level. In this work we assume a statistic with a *t-distribution*, for the limited number, $N_s$, of the surrogate realizations (typically $N_s$=10), giving a residual probability: p<0.01, for values of the statistic greather than 2.821 [15]. In principle, the Fourier-transform based surrogates method works very well but, in practice, there are many potential pitfalls and problems that can lead to false detection of nonlinearity [16]. Unlike analytical data, real signals are indeed contaminated with many external perturbations, such as *noise*, and also are measured with finite precision and recorded in finite sets. The situation is further complicated when the data exhibit *long coherence times*. Following Theiler et al., [16], we formalize the concept of coherence time in terms of the time $\tau$ such that the absolute value of the autocorrelation function is smaller than some pre-specified value $\varepsilon$ *for all t>$\tau$*. When a given time series exhibit long coherence times, the FT-based surrogates method, in spite of the theoretical expectation, does not correctly mimic all the linear properties of the original time series, and thus it can leads to uncorrect or false nonlinear statistic. In order to overcome this unwelcome feature, Paluš, [15], suggests to test differences between the original data and the related surrogates on *both the linear and nonlinear levels* to guarantee a reliable anonlinear statistic. In this work, to avoid spurious results due to linear deviations in the ensemble of surrogate data, we initially performed a *linear redundancy analysis*, both on the original data and on the surrogates ensemble. Following Paluš, we define the linear redundancy of an arbitrary n-dimensional random variable: $x_1,...,x_n$, by the formula:

$$L(x_1,...,x_n) := \frac{1}{2}\sum_{i=1}^{n} \log(c_{ii}) - \frac{1}{2}\sum_{i=1}^{n} \log(\sigma_i) \quad (8)$$

where $c_{ii}$ and $\sigma_i$ are the diagonal elements (variances) and the eigenvalues, respectively, of the *covariance matrix*, *C*, describing the mutual linear dependencies of the variables [15]. Through the computation of linear redundancies (8), we was able to select the *optimal* generic realization of the original data that guarantees the *linear compatibility* with the related ensemble of surrogates, and thus the reliability of results from the nonlinear statistic analysis.

Practical applications of the above nonlinear statistic analysis, often need reliable estimations of mathematical quantities, like correlation dimensions, that are strongly affected by the presence of broadband noise [11],[13]. The aim is the maximum reduction of the contaminating noise sources, without a significant distorsion of the original signal. In this work we further enhance the effectiveness of the filtering procedure successfully used in a previous study, [13], extending from the linear to the nonlinear framework, the class of filters proposed by Schreiber and Grassberger (1991) [17]. In this case the recurrent transformation for the filtered values, $x_n^*$, is expressed in the embedding space:

$$x_n^* = \sum_{i=-k}^{k} a_i^{(n)} x_{n+i} + b^{(n)} \qquad (9)$$

As in the linear case, the optimal values of the coefficients, $a_i^{(n)}$, $b^{(n)}$, are obtained by a least squares procedure restricted in a neighborhood of the embedding vector to be corrected [13],[17].

The whole procedure for detecting nonlinearity in real data adopted in the present work may be summarized in the following:

- Construction of an experimental time series from $N$ measurements of a selected representative scalar quantity (e.g. absolute pressure data);

- Partitioning of the original data in a set of $k$ sub-series of a given lenght $n$ ($=2^i \geq 512$): $nk \leq N$;

- For each sub-series, $i_k=1,2,...,k$, we generate $s$ (~10) surrogates and we check, using a linear redundancy analysis, if the linear characteristics of the original data match those of the surrogates ensemble;

- For each current collection: original sub-series+related surrogates, we compute the Takens best estimator of correlation dimension, $Q$, over a range of different embedding dimensions: $m_1, m_2, ..., m_s$ (here ~9): $Q^{(0)}(i_k, m_i)$, $Q^{(1)}(i_k, m_i)$,..., $Q^{(s)}(i_k, m_i)$; $i_k=1,2,...,k$, $m_i=m_1,...,m_s$;

- Estimation of the significance of nonlinear statistic to measure how significantly the original data differ from the related surrogates ensemble (see (7)): $S(i_k, m_i)=f(Q^{(\alpha)}(i_k, m_i))$; $\alpha=0,1,...,s$, $i_k=1,...k$, $m_i=m_1,...,m_s$;

- Evaluation of the mean value, computed from the k original subseries, of the sigmas, $S$, for each embedding dimension: $\hat{S}(m_i)=<S(i_k,m_i)>_{\{ik\}}$, $m_i=m_1,...,m_s$;

- Evaluation of the nonlinear degree for the original series computed, for example, through the mean value from the embedding dimensions: $\mathbb{N}_i:=<\hat{S}(m_i)>_{\{mi\}}=\mu$.

Of course, if there are evidences of stochastic behaviour due to the presence of contaminating broadband noise in the original data, we apply a proper filtering procedure to *clean* the data. Moreover, the above analysis may be completed by an estimation of the principal invariant dynamical quantities, like Lyapunov exponents, Kolmogorov entropy, etc. for a better characterization of the observed process.

3. APPLICATION TO REAL DATA: EXPERIMENTAL MEASUREMENTS

The above described method has been applied to different time series generated by absolute pressure measurements taken inside the *NOVE* hydroelectric power plant of ENEL SpA (the Italian National Board of Electricity) [13]. This plant is equipped with a vertical Francis turbine for a maximum power output of 72 MW, with a maximum flow rate of 80 m$^3$/s. The runner size is 2.75 m, and its rotational speed is 230.8 rpm (3.84 Hz); whereas the draft tube is 21.6 m long. Figure 1 shows the selected location of the fluid (water) pressure transducers. In particular, a first set of measurements is related to the point labelled 7 in the above figure, just below the runner (sample PR180); whereas a second set of measurements is related to the point labelled 9, inside the divergent (sample PDB). The dynamic regime conditions of turbine, during the measurements, correspond to an intermediate electric power rate of 30 MW, where anomalous dynamic behaviours have been detected. Previous analyses, [13], showed a clear evidence of a deterministic chaotic behaviour for irregular oscillations of gas cavities inside the Francis turbine. The aim of the present work is to complete the cited analysis quantifying the degree of nonlinearity of the related dynamics. Moreover, to shed more light on these complex phenomena, we tried to compare the results of the nonlinear analysis with another set of new measurements, performed after some mechanical modifications have been included in order to reduce the principal dynamic instabilities of the turbine at partial loads. These modifications consist in a conic-cylindrical appendix constrained to the final edge of the turbine shaft (suffix OFF), and an air flow inlet located in specific positions on the turbine shaft (suffix ON) (see figure 2). The original analogical electric signals, recorded on high speed magnetic tapes by ENEL SpA (DPT-SMP - Venice), have been converted and recorded in a suitable digital format through a special device available at CISE laboratories.

## 4. APPLICATION TO REAL DATA: ANALYSIS OF RESULTS

Nonlinearity analysis of time series: PR180 (test series), already explored in a previous work [13], performed on different filtered data (the original series, indeed, showed a typical random, somewhat correlated, behaviour due to a distributed noise), shows an *increase of the coherence time due to the filtering procedure*. This feature is well pointed out by the linear redundancy analysis, and it forces to a proper choice of the original sub-series lenght to avoid spurious detection of nonlinearity (figures 3,4,5). The nonlinear statistic, eq.(4), on the unfiltered data, clearly shows that the underlying dynamics is compatible with a linear stochastic process, with a strong evidence of a noise contamination [13],[16]. Moreover, the significance (7) sigmas suggests, in this case, a very low mean value of the nonlinearity degree: $\mu=0.88\pm0.62$, where 95% is the assumed confidence interval (figures 6,7). A comparison with the residual probability threshold value ($p<0.01$), suggests the impossibility to reject the null hypothesis of a linear stochastic process. On the other hand, the application of a filtering procedure to the above real data changes both qualitatively and quantitatively the features of the nonlinear statistic. Now the sigmas are well beyond the critical threshold value, clearly showing the presence of a nonlinear deterministic behaviour: $\mu=24.33\pm8.94$ (figures 8,9). This first result supports some previous indications according to which the observed stochastic behaviour of the original unfiltered series is only apparent, mainly due to a broadband noise contamination.

Starting from the above result we sistematically applied the nonlinearity analysis to all the new time series, in the presence of the above cited mechanical modifications. Here sampling time was: *dt=0.02 s* (50 Hz) spanning a time interval of 43 seconds (2150 points). From the time evolution of the four signals (PR180 OFF, PDB OFF; PR180 ON, PDB ON), we note a more or less complex feature of data with a broadband frequency domain. A qualitative inspection of the original signals, afterwards supported by a quantitative Fourier analysis, suggests an increase of the dynamical complexity degree in the case of the air flow inlet configuration (suffix ON) (figures 10,11). This feature is also confirmed by the autocorrelation analysis, showing a decrease of the *autocorrelation time*, [18], or equivalently the *average cycle time*, [19], for the air flow inlet case (see table 1). As summarized in table 2, all the original unfiltered time series exhibit a behaviour *compatible with a linear stochastic process*. The linear redundancy analysis shows a good linear preserving property of the surrogates for sub-series lenghts of 512 points, indicating low coherence times. Moreover, the nonlinearity statistic analysis clearly shows no significant differences between original and surrogates data (figures 12,13).

After the application of the Schreiber-Grassberger nonlinear filter, eq.(9), the increased coherence times, evidenced by the linear redundancy analysis, forced us to extend the generic sub-series minimum lenght to 1024 points. For the filtered time series the nonlinear statistic shows a typical feature of nonlinear deterministic systems (i.e. a finite value saturation for the Takens best estimator of correlation dimension), *with significant differences between original and surrogates data* (figures 14,15). As shown in table 2, the significance sigmas are now well beyond the critical threshold value, allowing the safe rejection of the null hypothesis of a linear stochastic model. We note that in the case of series with the suffix ON (air flow inlet), the numerical estimation of the nonlinear degree is quite reliable; whereas in the case of the suffix OFF there is an high indetermination of the mean values inside the assumed 95% confidence interval. However, as a general indication, the above results suggest a decrease of the nonlinearity degree for an air flow inlet configuration, with possible increase of the dynamical complexity due to the presence of more significant stochastic components.

To complete our analysis we computed also some important invariant quantities, like the Kolmogorov Entropy, the Lyapunov spectrum, etc., and the related results are summarized in table 3 (see [13]). This information allows to shed more light on the dynamical complexity degree of the signals, and to point out its relation with the nonlinearity degree. In fact, the main result emerging from these computations is that the air flow inlet involves *an increase of the dynamical instability*, when the dynamics is initially a combination of nonlinear deterministic and linear stochastic contributions; whereas the air flow inlet *reduces the dynamical complexity*, when the dynamics is initially governed by only nonlinear deterministic components (figure 16).

5. CONCLUSIONS

We have presented some experimental applications of the method for testing nonlinearity in time series, utilizing linear redundancy and FT-based surrogates nonlinear statistic analyses. This work, combined with previous analyses, strongly support the suggested mathematical picture on the dynamical behaviour of gas cavities inside the draft tube of Francis turbines [1],[2],[13]. The basic idea is that the adequate and effective mathematical description of such physical systems must belong to the framework of nonlinear deterministic differential equation models, with a low dimensionality, rather than to a linear stochastic, with infinite degrees of freedom, ones. We hope that the results coming from these new approches could give some help to the advancement in the understanding of the complex nature of these real phenomena.

Acknowledgments
This work was supported by ENEL SpA CRIS (Milano, Italy), and carried out in the frame of a 1994 joint contract between ENEL SpA and CISE-Tecnologie Innovative.

REFERENCES

[1] Fanelli, M. and Sello, S., *"Complex and chaotic response of a nonlinear oscillator with an isothermal gas spring"*, Meccanica, 27, (1992).

[2] Fanelli, M. and Sello, S., *"Strange attractors and chaotic behaviour in nonlinear mathamatical models of a gas cavity"*, Hydrofractals '93 - International Conference on Fractals in Hydroscience, Ischia (Italy), October (1993).

[3] Papoulis, A., Probability, Random Variables and Stochastic Processes, McGraw-Hill, (1965).

[4] Takens, F., *"Detecting nonlinearities in stationary time series"*, Int. Jou. of Bif. and Chaos, 3,2, (1993).

[5] Pompe, B., *"On some entropy methods in data analysis"*, Chaos, Solitons and Fractals, 4,1, (1994).

[6] Theiler, J., Eubank, S., Longtin, A., Galdrikian, B. and Farmer, J.D., *"Testing for nonlinearity in time series: the method of surrogate data"*, Physica D, 58, (1992).

[7] Prickard, D. and Theiler, J., *"Generating surrogate data for time series with several simultaneously measured variable"*, Preprint March, (1994).

[8] Tsay, R.S., *"Model checking via parametric bootstraps in time series analysis"*, Appl. Statist., 41,1, (1992).

[9] Theiler, J. and Eubank, S., *"Don't bleach chaotic data"*, Preprint December, (1993).

[10] Grassberger, P. and Procaccia, I., *"Measuring the strangeness of strange attractors"*, Physica D, 9, (1983).

[11] Theiler, J., *"Estimating fractal dimension"*, J. Opt. Soc. Am. A, 7,6, (1990).

[12] Packard, N.H., Crutchfield, J.P., Farmer, J.D. and Shaw, R.S., *"Geometry from a time series"*, Phys. Rev. Lett., 45, (1980).

[13] Fanelli, M. and Sello, S., *"Experimental time series analysis of dinamically irregular gas cavities in a real Francis turbine"*, XVII IAHR Symposium, 15-19 September, Beijing, China, (1994).

[14] Bendat, J.S. and Piersol, A.G., Random Data: Analysis and Measurement Procedures, Wiley-Interscience, (1971).

[15] Paluš, M., *"Testing for nonlinearity using redundancies: quantitative and qualitative aspects"*, Preprint, June (1994).

[16] Theiler, J., Linsay, P.S. and Rubin, D.M., *"Detecting nonlinearity in data with long coherence times"*, Time Series Prediction: Forecasting the Future and Understanding the Past, A.S. Weigend and N.A. Gershenfeld, Proc.Vol. XVII, Addison-Wesley, (1993).

[17] Schreiber, T. And Grassberger, P., *"A simple noise-reduction method for real data"*, Phys. Lett. A., 160, (1991).

[18] Theiler, J., *"Spurious dimension from correlation algorithms applied to limited time series data"*, Phys. Rev. A., 34,3, (1986).

[19] Schouten, J.C., Takens, F. and van den Bleek, C.M., *"Estimation of the dimension of a noisy attractor"*, Preprint Phys. Rev. E, February, (1994).

| Series | Dt | Tac | Cycle Time | Min. Minf. (1st zero Ac) Dt | Eff. Tac | Window | Points Window |
|---|---|---|---|---|---|---|---|
| PR180 OFF | 0.02 | 0.253 | 0.443 | 18 (16) | 0.25 | 0.44 | 22 |
| PR180 ON | 0.02 | 0.107 | 0.236 | 13 (12) | ---- | ---- | -- |
| PDB OFF | 0.02 | 0.050 | 0.155 | 20 (16) | 0.05 | 0.155 | 8 |
| PDB ON | 0.02 | 0.042 | 0.146 | -- (40) | ---- | ---- | -- |
| | | | | | | | |
| PR180 OFF F | 0.02 [3b] | 2.15 | 0.95 | 20 (18) | 2.07 | 1.23 | 31 |
| PR180 ON F | 0.02 [8] | 0.88 | 0.81 | 14 (15) | 0.88 | 0.8 | 40 |
| PDB OFF F | 0.02 [6] | 0.36 | 0.89 | 25 (23) | 0.38 | 1.48 | 37 |
| PDB ON F | 0.02 [7] | 0.45 | 0.83 | 20 (43) | 0.45 | 0.83 | 41 |

TABLE 1

Legenda: Series=Coded Name of Experimental Data; Dt=Original Time Sampling (s); Tac=Autocorrelation Time; Cycle Time=Average Cycle Time; Min. Minf.=1st Local Minimum of Mutual Information in Dt units; Ac=Autocorrelation Function; Window=Assumed Window Length from Sampled Time Series; F=Nonlinear Filtering Process of data.

| Series | Red.Opt. Ser.Length | Takens Best Est. | Nonlinear Degree (Sigmas) | 95% Conf. Level |
|---|---|---|---|---|
| PR180 OFF | 512 | ∞ | 2.25 | 0.67 |
| PR180 ON | 512 | ∞ | 1.08 | 0.45 |
| PDB OFF | 512 | ∞ | 2.46 | 2.8 |
| PDB ON | 512 | ∞ | 1.74 | 1.17 |
| | | | | |
| PR180 OFF F | 1024 | 13 | 20.5 | 4.25 |
| PR180 ON F | 512 | 10 | 16.7 | 1.9 |
| PDB OFF F | 1024 | 3 | 33.1 | 8.3 |
| PDB ON F | 512 | 5 | 15.85 | 2.89 |

TABLE 2

Legenda: Series=Coded Name of Experimental Data; Red. Opt. Ser. Length=Optimal Length of Subseries from Linearized Redundancy Analysis; Takens Best Est.=Dimension Estimation from Best Takens Estimator; Nonlinear Degree (Sigmas)= Nonlinearity Degree from Surrogate Series Analysis.

| Series | Limit Cut-Off W∞ | Max. Cut-Off Used Wm | Corr. Dim. | Emb. Dim. | Max. Lyap. Exponent | Lyap. Dim. | Kolm. Entropy |
|---|---|---|---|---|---|---|---|
| PR180 OFF | 13 | 3 | ---- | ---- | ---- | ---- | ---- |
| PR180 ON | ---- | ---- | ---- | ---- | ---- | ---- | ---- |
| PDB OFF | 3 | ---- | ---- | ---- | ---- | ---- | ---- |
| PDB ON | ---- | ---- | ---- | ---- | ---- | ---- | ---- |
| | | | | | | | |
| PR180 OFF F | 52 | 13 | 3.1 | 5 | 2.23 | 2.9 | 2.23 |
| PR180 ON F | 44 | 10 | 6.1 | 7 | 22.0 | 6.58 | 38.4 |
| PDB OFF F | 10 | 3 | 4.3 | 7 | 10.9 | 6.53 | 20.2 |
| PDB ON F | 23 | 5 | 4.6 | 7 | 4.0 | 3.5 | 4.0 |

TABLE 3

Legenda: Series=Coded Name of Experimental Data; Corr. Dim.=Correlation Dimension Estimation from Grassberger-Procaccia Method; Lyap. Dim.=Lyapunov Dimension Estimation from Kaplan-Yorke Congecture; Kolm. Entropy=Sum of the positive Lyapunov Exponents (Pesin).

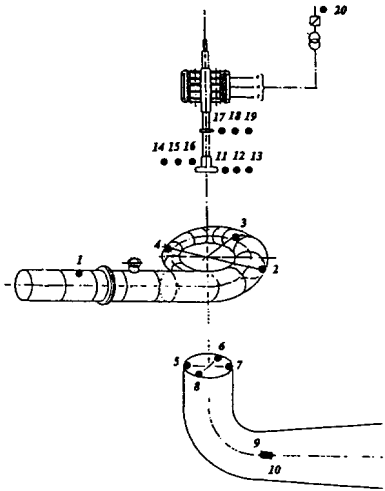

Figure 1
Selected measurement positions

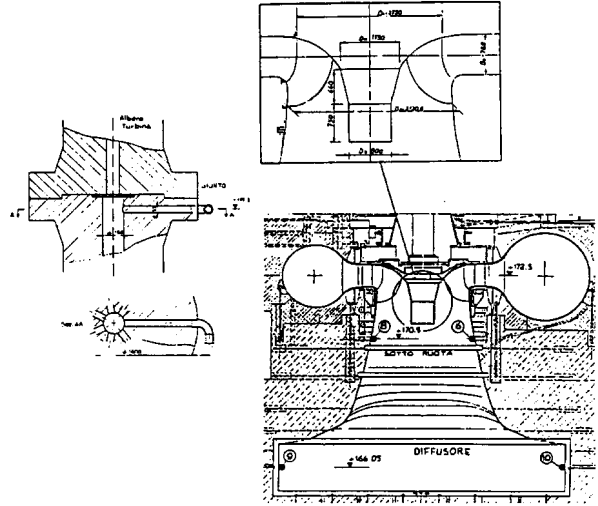

Figure 2
Mechanical modifications on the turbine shaft

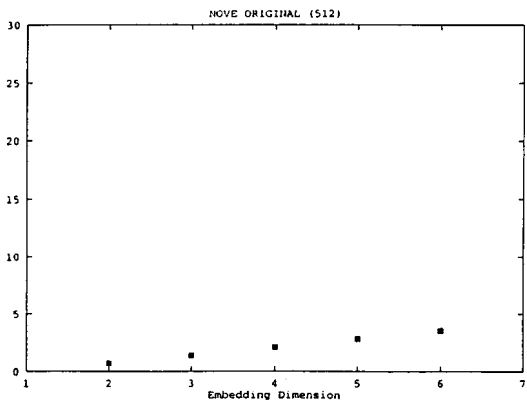

Figure 3
linear redundancy analysis: original data

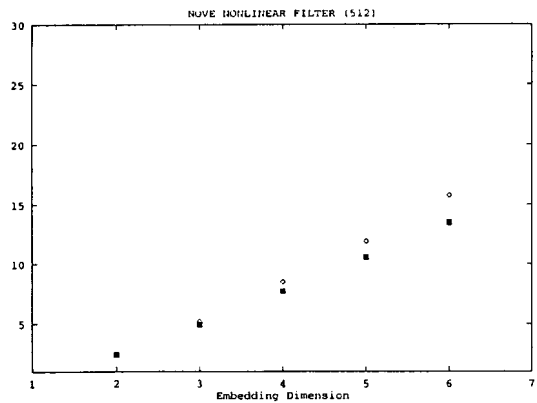

Figure 4
Linear redundancy analysis: filtered data

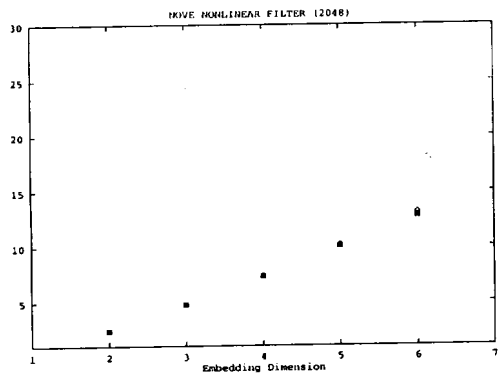

Figure 5  Redundancy Analysis

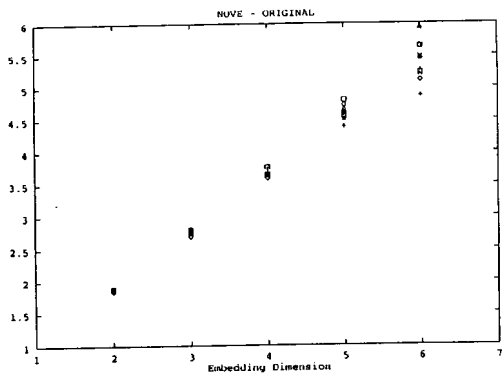

Figure 6  Takens Best Estimator

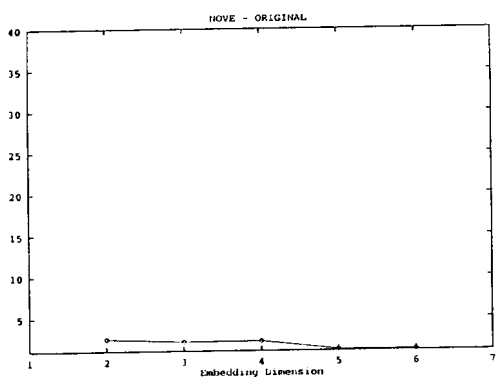

Figure 7  Significance (sigmas)

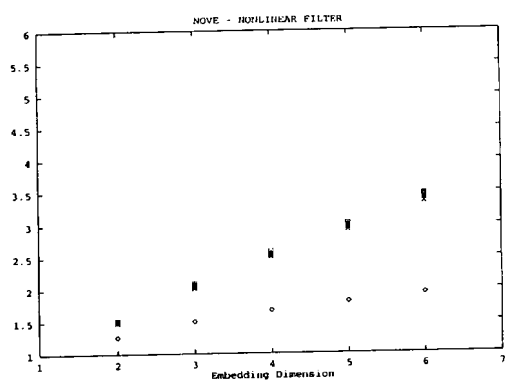

Figure 8  Takens Best Estimator

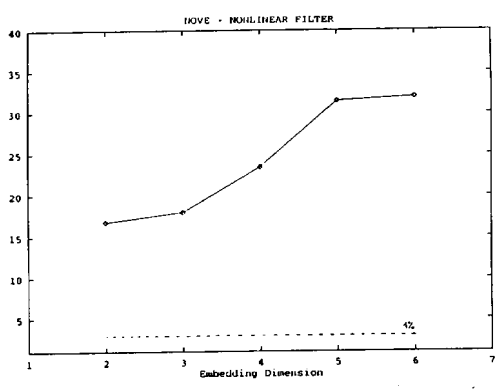

Figure 9  Significance (sigmas)

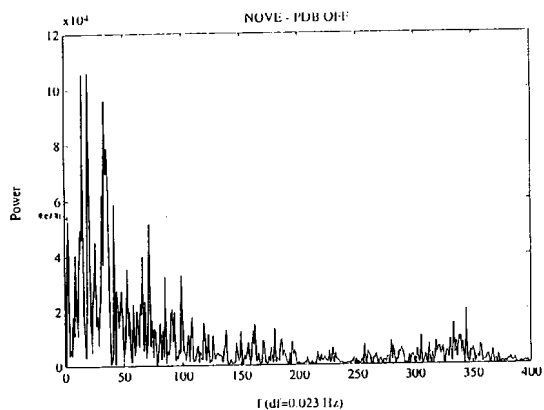

Figure 10  Fourier Analysis

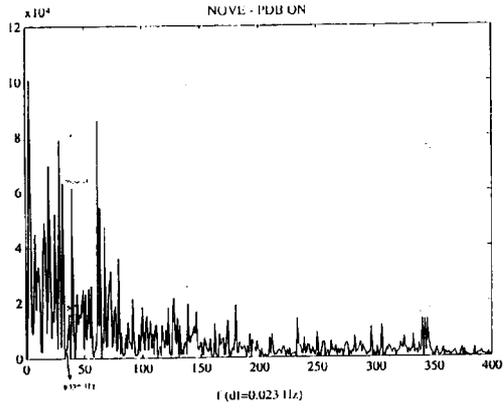

Figure 11 Fourier Analysis

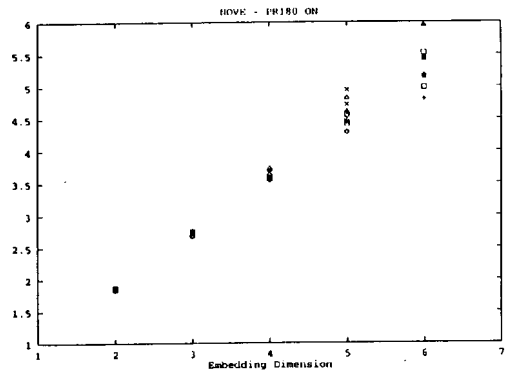

Figure 12 Takens Best Estimator

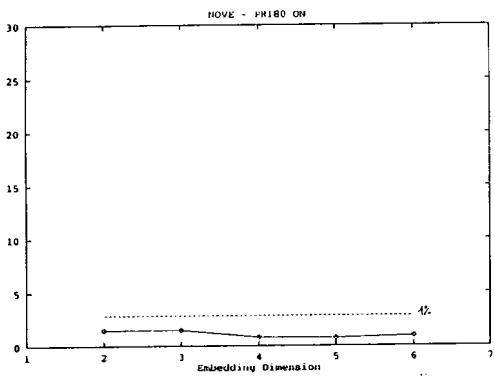

Figure 13 Significance (sigmas)

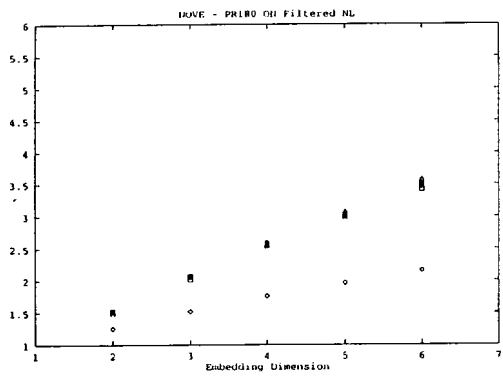

Figure 14 Takens Best Estimator

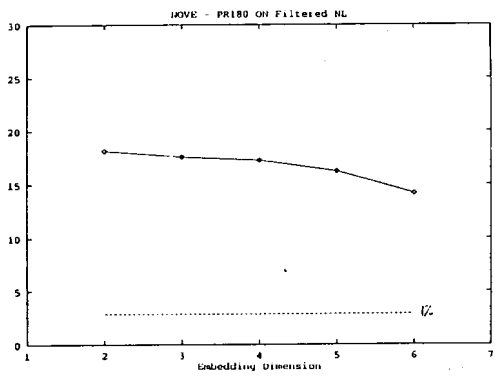

Figure 15 Significance (sigmas)

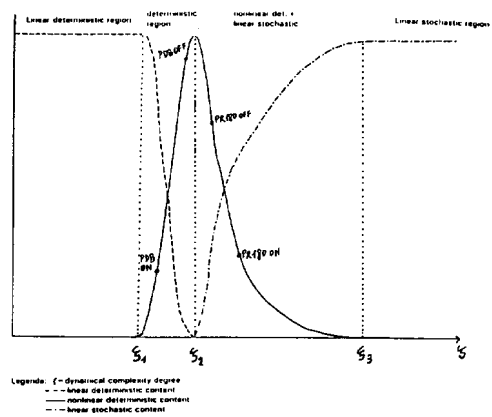

Figure 16 Nonlinearity/Complexity Model